\documentclass[
superscriptaddress,
nofootinbib,
twocolumn,
amsmath,amssymb,
aps,
prl,
]{revtex4-2}

\usepackage{amssymb}

\usepackage{bm}
\usepackage{mathrsfs}         
\usepackage{color}
\newcommand{\rev}[1]{{#1}}

\begin{document}

	\title{Classification of emergent Weyl spinors in multi - fermion systems}

	\author{M.A. Zubkov \footnote{On leave of absence from NRC "Kurchatov Institute" - ITEP, B. Cheremushkinskaya 25, Moscow, 117259, Russia}}
	
	\affiliation{Physics Department, Ariel University, Ariel 40700, Israel} 
	\affiliation{{\it On leave of absence from} NRC "Kurchatov Institute" - ITEP, B. Cheremushkinskaya 25, Moscow, 117259, Russia}
	
	\vspace{10pt}

	\begin{abstract}
		In the fermionic systems with topologically stable Fermi points the emergent two - component Weyl fermions appear. We propose the topological classification of these fermions based on the two invariants composed of the two - component Green function. We define these invariants using Wigner - Weyl formalism also in case of essentially non - homogeneous systems. In the case when values of these invariants are minimal ($\pm 1$) we deal with emergent relativistic symmetry. The emergent gravity appears, and our classification of Weyl fermions gives rise to the classification of vierbein. Transformations between emergent relativistic Weyl fermions of different types correspond to parity conjugation, time reversal, and charge conjugation.
	\end{abstract}
	
	%
	%
	%
	%
	%





	\maketitle

\section{Introduction}

Electrons in solids are described by multi - component spinors carrying band index. However, at low energies we may describe electrons by effective spinors with the essentially reduced number of components. Only those energy bands are relevant that cross Fermi energy. In Dirac/Weyl semimetals \cite{Abrikosov1971,Abrikosov1998,Burkov2011,Burkov2011_,XiangangWan2011,Weylsemimetal} with Fermi points the emergent spinors are two component if Fermi energy is close to the position of the Fermi points. Due to the repulsion of energy levels the Fermi points are unstable unless they are protected by topology. Therefore, the effective description in terms of the two - component spinors typically survives in the case when the topological invariants protecting the Fermi points are nonzero \cite{Horava2005}. The earlier discussion of these issues may be found in  \cite{Novikov1981,Avron1983,Volovik1987}. For the review of the topological invariants (in momentum space) protecting Fermi points and Fermi surfaces see \cite{VolovikBook}. The well - known example is graphene with emergent Weyl fermions \cite{Semenoff:1984dq,CastroNeto:2009zz}. It is also worth mentioning that the topological invariants are responsible for gapless edge modes of
topological insulators
\cite{VolkovPankratov1985,HasanKane2010,Xiao-LiangQi2011}. The similar phenomena are observed also in superfuids
\cite{SalomaaVolovik1988,Volovik2009}.

The minimal value of the topological invariant $N_3$ (composed of the Green functions \cite{VolovikBook}) responsible for the stability of the Weyl points is $\pm 1$. Weyl points with the larger values of $N_3$ are able to split into several Weyl points with $N_3=\pm 1$. Action of Weyl fermions with $N_3 = \pm 1$ is relativistic. This results in the appearance of emergent vierbein
\cite{Froggatt1991,VolovikBook,Volovik1986A,Volovik2011}. The emergent Weyl fermions move effectively in a curved space - time. The emergent gravity fluctuates, but these fluctuations are not described by a theory invariant under the diffeomorphisms.

The nonminimal values of $N_3$ give rise to more exotic types of  Weyl fermions, typically with the nonlinear touching
points of positive and negative energy branches. An example of such exotic Weyl points is given by the $2+1$ D multilayer graphene  with
the $ABC$ stacking
\cite{multilayer}. Also such exotic Weyl fermions appear in effective
gravitational theories with anisotropic scaling
\cite{HoravaPRL2009,HoravaPRD2009,Horava2008,Horava2010}, see
\cite{KatsnelsonVolovik2012,KatsnelsonVolovikZubkov2013a,Zubkov2012,KatsnelsonVolovikZubkov2013b,VZ2013gr,VZ2013gr_}.

It is worth mentioning that the classification of topological insulators
\cite{Schnyder2008,Schnyder2008_,Kitaev2009} may be obtained by a certain reduction of the topological classification of Fermi surfaces and emergent spinors incident on them \cite{Horava2005}, for the details see \cite{VolovikBook}.

In the present paper we are going to extend the mentioned above classification of Weyl points into two directions. First of all, we notice that in general case of interacting systems the two topological invariants $N_3$ and $N_3^{(3)}$ may be introduced. Both are composed of the Green functions, and both of them are reduced to the above mentioned $N_3$ in the absence of interactions. However, they become different in general case. As a result in case of the minimal values $N_3,N_3^{(3)} = \pm 1$ we obtain the four topologically distinct types of Weyl fermions. We call two of them the left - handed and the right handed particles, and the other two - the left - handed and the right - handed "anti - particles". The latter types of Weyl points are referred to as the anti - Weyl points, they may appear only in the presence of interactions. This classification extends the conventional one, which considers the two types of relativistic Weyl fermions - the left - handed, and the right - handed. Such an extension was noticed, in particular, in \cite{Volovik2020}, though without reference to appropriate topological invariants. Transformations between topologically distinct types of emergent relativistic Weyl fermions are given by various combinations of charge conjugation, time reversal, and parity conjugation. These transformations lead to the specific transformation of vierbein (see \cite{Vergeles2019}).

Another direction for the extension of the classification of emergent Weyl fermions is related to consideration of the non - homogeneous systems following methodology of \cite{ZW2019,ZZ2019,SZ2020}. Namely, the conventional expression for invariant $N_3$ is defined as an integral in momentum space and is valid for the homogeneous systems only. Using the Wigner - Weyl formalism we extend the expressions of both $N_3$ and $N_3^{(3)}$ to the non - homogeneous case, when they are given by integrals in phase space.

\section{Emergent Weyl spinors in the multi - fermion systems}

\label{sectHorava}

Here we closely follow \cite{Volovik:2014eca}. We start from the consideration of  equilibrium condensed matter system with
the $n$ - component spinors $\psi$ at zero temperature. Its real time partition function has the form:
\begin{eqnarray}
Z & = & \int D \psi D \bar{\psi} {\rm exp}\Bigl(i\int d t \sum_{{\bf x}}
\bar{\psi}_{\bf x}(t) (i\partial_t \nonumber\\&&- \hat{H}_{\rm }e^{- i \epsilon }) \psi_{\bf x}(t)
\Bigr),\label{ZH}
\end{eqnarray}
Hamiltonian $H$ is a Hermitian matrix depending on momentum
$\hat{\bf p}$. Sum over points of coordinate space is to be understood as an integral over $d^3 x$ for continuous coordinate space. However, we may also consider lattice tight - binding models, in which case we have the sum over the lattice points.
In the absence of dependence of $H$ on $x$ Hamiltonian $H$ depends on $\bf p$ only. Factor $e^{- i \epsilon }$ is introduced here in order to point out how the poles in Feynmann diagrams are to be treated. $\epsilon$ is assumed to be very small, its appearance may be explained, for example, via equilibrium limit of the Keldysh non - equilibrium path integral in lattice regularization (of time axis).  Typically the factor $e^{-i\epsilon }$ is omitted and is restored only if it is necessary to speak about the ways the singularities in Feynmann diagrams are treated.

Several branches of spectrum of $\hat H$ repel
each other. Therefore, the minimal number $n_{\rm
reduced}=2$ of branches are able to cross each other. And this minimal number is fixed by topology of momentum
space.

Let us consider the position ${\bf p}^{(0)}$ of the crossing of 2 branches of $\hat H$. Transformation via a certain Hermitian matrix $\Omega$:
 $\tilde{H}( {\bf p}) = \Omega^+ \hat{H} \Omega$ brings Hamiltonian to the diagonal form. Then the first $2 \times 2$
block $\hat{H}_{\rm reduced}$ corresponds to the two crossed branches. The remaining block $\hat{H}_{\rm gapped}$
corresponds to the branches of spectrum separated from the crossing point by a gap. We may represent the functional integral as the product of the functional integral over "gapped" modes $\Theta$
and the integral over 2 reduced fermion components $\Psi$. At low energies (close to the Fermi level coinciding with the branch crossing point) the contribution to the physical observables of the reduced fermions dominates over the contribution of the gapped ones because $\Theta$ contribute
the physical quantities with the fast oscillating factors. Therefore, we effectively describe the system close to the Fermi point via the partition function
\begin{eqnarray}
Z &=& \int D \Psi D\bar{\Psi}  {\rm exp}\Big(i\int d t \sum_{{\bf x}}
\bar{\Psi}_{\bf x}(t) (i\partial_t \nonumber\\&& - \hat{H}_{\rm reduced}e^{-i\epsilon}) \Psi_{\bf x}(t)
\Big)\label{ZH3}
\end{eqnarray}
In the following we will omit the subscript "reduced" of $\hat{H}$ for simplicity.
The general form of Hermitian matrix $2\times 2$ brings the partition function to the form
\begin{eqnarray}
Z &=& \int D \Psi D \bar{\Psi} {\rm exp}\Bigl(  i \int d t \sum_{{\bf x}}
\bar{\Psi}_{{\bf x}}(t) (i\partial_t \nonumber\\&& - [m^L_k(\hat{\bf p}) \hat \sigma^k  +
m(\hat{\bf p})]e^{-i \epsilon })\Psi_{\bf x}(t) \Bigr)\label{ZH__0}
\end{eqnarray}
where functions $m^L_k, m$ are real - valued. The nontrivial topology appears when the topological invariant composed of ${ m}^{L}({\bf p})$ has the nontrivial value
\begin{equation}
N_3= \rev{-}\frac{e_{ijk}}{8\pi} ~
   \int_{\sigma}    dS^i
~\hat{ m}^L\cdot \left(\frac{\partial \hat{ m}^L}{\partial p_j}
\times \frac{\partial \hat{ m}^L}{\partial p_k} \right), \quad \hat{ m}^L =
\frac{{ m}^L}{|{ m}^L|}
\label{NH}
\end{equation}
where $\sigma$  is the $S^2$ surface surrounding the branch crossing point $p^{(0)}_j$.

On the language of the Matsubara Green function
 $$
 G_H(\omega, p) = \frac{1}{i\omega - \hat{H}({\bf p})}
 $$
 the value of $N$ may be written as ($\omega = p_4$):
\begin{eqnarray}
N_3^{(3)}  &=&     \frac{1}{3!\,4\pi^2} \int_\Sigma \, {\rm Tr}  \, \Big[  {G}_H(\omega, {\bf p}) d {G}^{-1}_H(\omega, {\bf p})\nonumber\\&&  \wedge d {G}_H( \omega, {\bf p}) \wedge  d {G}^{-1}_H(\omega, {\bf p}) \Big]\label{rez2}
\end{eqnarray}
Here $\Sigma$ is the three - dimensional hypersurface surrounding the Fermi point in $4D$ momentum space (composed of $3D$ momentum space and the axis of $\omega = p_4$).

If $N=\pm 1$ in Eq.(\ref{NH}), then close to $p = p^{(0)}$ the following expansion takes place
\begin{equation}
m^L_a({\bf p})=f_a^j(p_j-p^{(0)}_j)\,.
\label{A(K)-expansion0H}
\end{equation}
with ${\rm sign}\, {\rm det} f_a^j = N$,
and
$$
m({\bf p}) \approx f_0^j(p_j-p^{(0)}_j )
$$
with  $j =
1,2,3$.
The coefficients $f^j_a$ ($a=0,1,2,3$) of expansion may be considered as the source of emergent gravity.
As a result, Eq. (\ref{ZH__0}) receives the form ($\sigma^0 \equiv 1$):
\begin{eqnarray}
Z &=& \int D \Psi D\bar{\Psi} {\rm exp}\Bigl(i\int d t \sum_{{\bf x}}
\bar{\Psi}_{{\bf x}}(t) (i\partial_t \nonumber\\ && - f_a^j(\hat{\bf p}_j-p^{(0)}_j) \hat
\sigma^a e^{-i \epsilon } )\Psi_{\bf x}(t) \Bigr)\label{ZHH}
\end{eqnarray}
The dispersion of quasiparticles close to the point $p^{(0)}$ receives the form of the Dirac cone. Due to nonzero values of $f_0^k$ this cone is tilted. Moreover, if $\|f^j_0 [f^{-1}]_j^a\| > 1$ then the cone becomes over - tilted and we come to the concept of the so - called type II Weyl point \cite{typeII} (see Remark 2.1. in \cite{Volovik:2014eca}, where this kind of Weyl point has been proposed for the first time), when close to the Fermi point the two Fermi pockets appear. The form of the latter is given by solution of equation
\begin{equation}
{\rm det} \, \sigma^a f_a^j(p_j-p^{(0)}_j) = 0, \quad j = 1,2,3; \, a=0,1,2,3
\label{cfs}
\end{equation}
The intermediate state between the ordinary (type I) Weyl point and the type II Weyl point corresponds to the case, when $\|f^j_0 [f^{-1}]_j^a\| = 1$. In this case the Fermi point (the couple of the Fermi pockets) is reduced actually to the Fermi line.

\section{Topological invariants and emergent vierbein in homogeneous systems}
\label{sectinter0}

In the presence of interactions the
partition function can be written as follows:
\begin{eqnarray}
Z &=& \int D \psi D\bar{\psi} D\Phi {\rm exp}\Bigl(i R[\Phi] + i \int d t
\sum_{{\bf x}} \bar{\psi}_{\bf x}(t) (i\partial_t \nonumber\\&&- \hat{H}(\Phi)e^{-i\epsilon }) \psi_{\bf
x}(t) \Bigr)\label{FIn}
\end{eqnarray}
Here $\Phi$ is a set of fields that provide interactions. $R$ is a certain effective action of $\Phi$. Operator $\hat
H$ also depends on $\Phi$. In mean field approximation we substitute the values of $\Phi$ by their "mean" values. Fluctuations around these mean values will give us the partition function of the form
\begin{eqnarray}
Z &=& \int D \Psi D \bar{\Psi} D\Phi  \nonumber\\&&{\rm exp}\Bigl(i R[\Phi]\Bigr) {\rm
exp}\Bigl(  i \int d t \sum_{{\bf x}} \bar{\Psi}_{{\bf x}}(t) (i\partial_t m^\prime_{\Phi,a} \sigma^a  \nonumber\\&& + [\mu m^\prime_{\Phi,a} \sigma^a -
m^L_{\Phi, k}(\hat{\bf p}) \hat \sigma^k  - m_{\Phi}(\hat{\bf p})]e^{-i\epsilon })\Psi_{\bf
x}(t) \Bigr)\label{ZH__0}
\end{eqnarray}
Here it is assumed that $\Psi$ carries an index enumerating the Weyl points of the system.  

Let us consider the Green function of the original multi - fermion system
 \begin{eqnarray}
&& G(t_1 - t_2, x-y) =  \frac{-i}{Z}   \int D \psi D\bar{\psi} D\Phi {\rm exp}\Bigl(i R[\Phi]  \nonumber\\&&+ i \int d t
\sum_{{\bf x}} \bar{\psi}_{\bf x}(t) (i\partial_t   - \hat{H}(\Phi)e^{-i \epsilon}) \psi_{\bf
x}(t) \Bigr) \nonumber\\&& \psi_x(t_1) \bar{\psi}_y(t_2)\label{Gc}
\end{eqnarray}
 In order to construct the topological invariants responsible for the stability of Fermi points we may also use the Green function of the reduced low energy theory
  \begin{eqnarray}
&& G(t_1 - t_2, x-y) =  \frac{-i}{Z}   \int D \Psi D \bar{\Psi} D\Phi  \nonumber\\&&{\rm exp}\Bigl(i R[\Phi]\Bigr) {\rm
exp}\Bigl(  i \int d t \sum_{{\bf x}} \bar{\Psi}_{{\bf x}}(t) (i\partial_t m^\prime_{\Phi,a} \sigma^a  \nonumber\\&& + [\mu m^\prime_{\Phi,a} \sigma^a -
m^L_{\Phi, k}(\hat{\bf p}) \hat \sigma^k  - m_{\Phi}(\hat{\bf p})]e^{-i\epsilon })\Psi_{\bf
x}(t) \Bigr) \nonumber\\&& \Psi_x(t_1) \bar{\Psi}_y(t_2)\label{Gc_}
\end{eqnarray}
Then we compose the  Fourier transform $G(P_0,P_1,P_2,P_3)$ of the Green function either from Eq. (\ref{Gc}) or from Eq. (\ref{Gc_}). We substitute to $G(P_0,P_1,P_2,P_3)$ the values $P_0 = i \omega$ and $P_j = p{_j}$ for $j = 1,2,3$. 

 The first topological invariant is given by
\begin{eqnarray}
N_3  &=&     \frac{1}{3!\,4\pi^2} \int_\Sigma \, {\rm Tr}  \, \Big[  {G}(\omega, {\bf p}) d {G}^{-1}(\omega, {\bf p}) \nonumber\\&&  \wedge d {G}_H(\omega, {\bf p}) \wedge  d {G}^{-1}_H(\omega, {\bf p}) \Big]\label{rez3}
\end{eqnarray}
Here $\Sigma$ is the three - dimensional hypersurface surrounding the Fermi point in $4D$ momentum space (composed of $3D$ momentum space and the axis of $\omega = p_4$). 

In addition, we define
 \begin{equation}
 G_H(\omega, {\bf p}) = \frac{1}{i\omega - G(0, {\bf p})^{-1}}\label{G_H}
 \end{equation}
(Again, we may use either G of Eq. (\ref{Gc}) or Eq. (\ref{Gc_}).)
The second topological invariant is defined as
\begin{eqnarray}
N_3^{(3)}  &=&     \frac{1}{3!\,4\pi^2} \int_\Sigma \, {\rm Tr}  \, \Big[  {G}_H(\omega, {\bf p}) d {G}^{-1}_H(\omega, {\bf p})  \nonumber\\&& \wedge d {G}_H( \omega, {\bf p}) \wedge  d {G}^{-1}_H(\omega, {\bf p}) \Big]\label{rez2}
\end{eqnarray}

For the minimal values of $N_3$ and $N_3^{(3)}$ at each Weyl point we represent
\begin{eqnarray}
&& m^{\prime}_{\Phi, a}({\bf p})\approx e \, e_a^0,  \quad
m^{L}_{\Phi, a}({\bf p})\approx e \, e_a^j({p}_j - B_j ),   \nonumber\\&&
m_{\Phi}({\bf p}) \approx  e\, e_0^j({p}_j-B_j ),\quad i,j = 1,2,3
\label{A(K)-expansion0120}
\end{eqnarray}
Here we assume that fields $\Phi$ as well as $e_\mu^a$ are slowly varying. 
(Again, an index referring to the number of the Fermi point is suppressed.)
We introduce here notations that match relativistic quantum field theory. By $e$ we denote the determinant of matrix inverse to the $4\time 4$ matrix $e_a^k$.
The appearance of the field $B_0 = -\mu$ reflects, that in the presence of
interaction the value of Fermi energy at the position of the crossing of several branches of spectrum may differ from zero. Thus, $\mu$ is chemical potential counted from the crossing point.

As a result, the partition function of the model may be rewritten as:
\begin{equation}
Z = \int D \Psi D\bar{\Psi} D e^i_k D B_k e^{i S[e^j_a, B_j,
\bar{\Psi},\Psi]}\label{ZI}
\end{equation}
with
\begin{eqnarray}
&& S =   S_0[e, B]  \nonumber\\&&+ \frac{1}{2} \Bigl(i \,\int d t \,e\,  \sum_{{\bf x}}
\bar{\Psi}_{{\bf x}}(t)  e_a^j  \hat \sigma^a  \hat D_j  e^{-i\epsilon} \Psi_{\bf x}(t) +
(h.c.)\Bigr),\label{Se0}
\end{eqnarray}
where the sum is over $a,j = 0,1,2,3$ while $\sigma^0 \equiv 1$, and $\hat
D$ is the covariant derivative that includes the $U(1)$ gauge field $B$:
$$
\hat{D}_0 = \partial_t e^{i\epsilon } - i \mu, \quad \hat{D}_k = \partial_k + i B_k
$$
$S_0[e, B]$ is the part of the effective action that depends on $e$ and $B$
only.

In the absence of an external source of inhomogeneity in the leading order the field $e_a^j$ may be considered as independent of coordinates. The same refers to $B$, which then may be gauged off. In this case in order to define topological invariants $N_3$ and $N_3^{(3)}$ we use the Green function
\begin{equation}
 G(\omega, {\bf p}) = \frac{1}{e\, \rev{(e_a^0  \hat \sigma^a  i \omega  - e_a^j  \hat \sigma^a p_j )}}\label{Gr}
\end{equation}
Obviously, $N_3^{(3)}$ and $N_3$ may differ from each other due to the nontrivial components of the emergent vierbein $e_a^0$. The direct calculation gives
$$
N_3 = \rev{-}{\rm sign}\, {\rm det}_{4\times 4} \, e_a^j
$$
(Here $a,j = 0,1,2,3$, and the determinant is of the $4\times 4 $ matrix.)
In the similar way we get
$$
N_3^{(3)} = \rev{-}{\rm sign}\, {\rm det}_{3\times 3}\, e_a^j
$$
where $a,j = 1,2,3$, and the determinant is of the $3\times 3$ matrix composed of the space components of the vierbein.

\section{T, C, P transformations and $N_3$, $N_3^{(3)}$.}

Let us consider the interplay of these topological invariants and T, C, P transformations. For the left - handed fermion (the one with $N_3 = N_3^{(3)}=+1$) the transformations are (we use matrix parts of those transformations, without transitions between the left - handed and the right handed spinors):
$$
T: \Psi(t,x) \to \sigma_2 \bar{\Psi}^T(-t,x), \quad \bar{\Psi}(t,x) \to \Psi^T(-t,x) \sigma_2
$$
$$
P: \Psi(t,x) \to -i  {\Psi}(t,-x), \quad \bar{\Psi}(t,x) \to \bar{\Psi}(t,-x)i
$$
$$
C: \Psi(t,x) \to \sigma_2 \bar{\Psi}^T(t,x), \quad \bar{\Psi}(t,x) \to \Psi^T(t,x)\sigma_2
$$
At the same time for the right - handed ones (those with $N_3 = N_3^{(3)}=-1$):
$$
T: \Psi(t,x) \to \sigma_2 \bar{\Psi}^T(-t,x), \quad \bar{\Psi}(t,x) \to \Psi^T(-t,x) \sigma_2
$$
$$
P: \Psi(t,x) \to -i  {\Psi}(t,-x), \quad \bar{\Psi}(t,x) \to \bar{\Psi}(t,-x)i
$$
$$
C: \Psi(t,x) \to -\sigma_2 \bar{\Psi}^T(t,x), \quad \bar{\Psi}(t,x) \to -\Psi^T(t,x)\sigma_2
$$
Let us also introduce the additional transformation R (both for the right - handed and for the left - handed fermions):
$$
R: \Psi(t,x) \to - \Psi(t,x), \quad \bar{\Psi}(t,x) \to \bar{\Psi}(t,x)
$$

These transformations lead to the same form of the action for the transformed spinor fields, in which the vierbein is transformed as follows:
\begin{eqnarray}
&& T:  e^0_0   \to  - e_0^0, \quad e^0_a \to e^0_a, \quad e^k_0 \to e_0^k , \quad e_a^k \to  - e_a^k,  \nonumber\\&&
 P:  e^0_0   \to   e_0^0, \quad e^0_a \to e^0_a, \quad e^k_0 \to -e_0^k , \quad e_a^k \to  -e_a^k,  \nonumber\\&&
 C:  e^0_0   \to   e_0^0, \quad e^0_a \to -e^0_a, \quad e^k_0 \to e_0^k , \quad e_a^k \to  - e_a^k,  \nonumber\\&& \quad a,k=1,2,3
\end{eqnarray}
One can check that the above transformations lead to the transformations of the topological invariants $N_3, N_3^{(3)}$:
$$
T: N_3 \to N_3, \quad N_3^{(3)} \to  -N_3^{(3)}
$$
$$
C: N_3 \to -N_3, \quad N_3^{(3)} \to  -N_3^{(3)}
$$
$$
P: N_3 \to -N_3, \quad N_3^{(3)} \to - N_3^{(3)}
$$
$$
CPT: N_3 \to N_3, \quad N_3^{(3)} \to  - N_3^{(3)}
$$
$$
R: N_3 \to N_3, \quad N_3^{(3)} \to  - N_3^{(3)}
$$

In the limit of small $\epsilon$ the operator $Q=e^k_a i D_k \sigma^a e^{-i \epsilon}$ is Hermitian. The same refers to operator $Q$ standing in the fermionic action $\int d^3x dt \bar{\Psi} Q \Psi$ also in general case. When $Q$ remains in the class of such (almost) Hermitian operators the smooth modifications of the system cannot lead to the change of the values of both $N_3$ and $N_3^{(3)}$.
 Different values of the topological invariants $N_3$ and $N_3^{(3)}$ determine the topological classification of Weyl fermions.
 Namely, we define those with $N_3 = N_3^{(3)} = +1$ as the left - handed particles. Notice, that our definition of $N_3$ differs by sign from that of \cite{VolovikBook}. The fermions with $N_3 = -N_3^{(3)}=1$ may be defined as the left - handed anti - particles. Those with $N_3 =  N_3^{(3)} = -1$ may be identified with the right - handed particles while $N_3 = - N_3^{(3)} = -1$ would correspond to the right - handed anti - particles. This is summarized in Table. One can see, that parity conjugation changes chirality of particles while time reversal transforms particles to anti - particles.
At the same time the RCPT (that includes $R:\Psi \to - \Psi$) is identical to unity, which reflects  invariance of fermionic systems under the CPT.

\begin{table}[ht]\caption{Weyl fermions and values of topological invariants.}
\centering 
\begin{tabular}{c c c }
\hline\hline                        
fermion type & $N_3$ & $N_3^{(3)}$ \\ [0.5ex]
\hline                  
left - handed Weyl point & +1 & +1 \\
right - handed Weyl point & -1 & -1   \\
left - handed anti - Weyl point & +1 & -1  \\
right - handed anti - Weyl point & -1 & +1  \\ [1ex]      
\hline 
\end{tabular}\label{table}
\end{table}

From the definition of the topological invariants $N_3$ and $N_3^{(3)}$ it follows that the sum of the values of $N_3$ or $N_3^{(3)}$ over all Fermi points is equal to zero provided that momentum space is compact. The proof of this statement is as follows. The sum of $N_3$ ($N_3^{(3)}$) over the Fermi points is given by Eq. (\ref{rez3}) (Eq.  (\ref{rez2})) with hypersurface $\Sigma$ surrounding all Fermi points. If total momentum space is compact, $\Sigma$ may be deformed smoothly to a point. The result is, obviously, zero. We deal with compact momentum space if lattice systems are considered with compact Brillouin zone, and, in addition, the axis of imaginary time is discretized. The latter condition is not always fulfilled, and we deal in condensed matter systems without interactions with the Green functions of the form of Eq. (\ref{G_H}). Then at $\omega \to \pm \infty$ the expression standing in the integrals of Eq. (\ref{rez2}) tends to zero. As a result the mentioned topological theorem is back, and the sum of $N_3^{(3)}$  over all Weyl points is zero. However, in the presence of interactions the dependence of the Green function on $\omega$ may be more complicated. Then the sum over the Weyl points of $N_3$ may appear to be nonzero while the sum of $N_3^{(3)}$ remains vanishing. In the latter case the weakened topological theorem allows breakdown of the conventional state with equal number of left and right - handed fermions in the lattice models. Namely, the situation is possible, when there are two \rev{right}  - handed fermions, but one of them is of the "particle" type, while another one is of the type of "anti - particles". This situation is considered in the next Section. 

\section{Toy model}

Let us consider the toy model, which illustrates unconventional properties of the systems with the anti - Weyl points (i.e. Fermi points corresponding to  the  "anti - particles"). In this model the partition function is given by
\begin{eqnarray}
Z &=& \int D \Psi D \bar{\Psi}  \nonumber\\&& {\rm
exp}\Bigl(  i \int d t \sum_{{\bf x}} \bar{\Psi}_{{\bf x}}(t) \hat{Q}(i\partial_t, \hat{\bf p}) \Psi_{\bf
x}(t) \Bigr)
\end{eqnarray}
Operator $\hat{Q}$ in momentum space has the form 
\begin{eqnarray}
&& \hat{Q}(p_0, {\bf p}) = p_0 \Big(\alpha \, {\rm cos}\,p_1 + (\alpha - 1) {\rm sin}\,p_1 \Big) \\ &&
- \Big(\alpha \, {\rm sin}\,p_1 - (\alpha - 1)( - {\rm cos}\,p_1 + {\rm cos}\, p_0 - 2) \Big)\sigma^1 e^{-i \epsilon} \nonumber\\&&
-{\rm sin} \,p_2 \sigma^2 e^{-i \epsilon} - ({\rm sin}\,p_3 + 2 - {\rm cos} p_2 - {\rm cos}\, p_3) \sigma^3 e^{-i \epsilon} \nonumber
\end{eqnarray}
Here $\alpha$ is a real  - valued  parameter.
In this system there are two Fermi points $K^\pm$:
$$
K^+: p_1 = 2 \, {\rm arctg}\, (1/\alpha - 1)=2\phi_\alpha, \quad p_2=p_3=0
$$
$$
K^-: p_1 = \pi, \quad p_2=p_3=0
$$
Operator $\hat{Q}$ may be written in the form:
\begin{eqnarray}
&& \hat{Q}(p_0, {\bf p}) = \sqrt{2\alpha^2 - 2\alpha +1} \Big(p_0 {\rm cos}\,(p_1 + \phi_\alpha ) \nonumber\\ &&
-  ( {\rm sin}\,(p_1 - \phi_\alpha) + {\rm sin}\,\phi_\alpha \,({\rm cos}\, p_0 -2))\sigma^1 e^{-i \epsilon} \nonumber\\&&
-\frac{{\rm sin} \,p_2 \sigma^2 e^{-i \epsilon} + ({\rm sin}\,p_3 + 2 - {\rm cos} p_2 - {\rm cos}\, p_3) \sigma^3 e^{-i \epsilon} }{\sqrt{2\alpha^2 - 2\alpha +1} }\Big)\nonumber
\end{eqnarray}
Close to the  Fermi points we have:
\begin{eqnarray}
K^+ & : & \frac{\hat{Q}(p_0, {\bf p})}{\sqrt{2\alpha^2 - 2\alpha +1}}  \nonumber\\ &=&  \Big(p_0 {\rm cos}\,\phi_\alpha
-  q_1  \, \sigma^1 e^{-i \epsilon}
-\frac{q_2 \sigma^2 e^{-i \epsilon} + q_3  \sigma^3 e^{-i \epsilon} }{\sqrt{2\alpha^2 - 2\alpha +1} }\Big)\nonumber
\end{eqnarray}
and
\begin{eqnarray}
K^- & : & \frac{\hat{Q}(p_0, {\bf p})}{\sqrt{2\alpha^2 - 2\alpha +1}}  \nonumber\\ &=&  \Big(- p_0 {\rm cos}\,\phi_\alpha
+  q_1  \, \sigma^1 e^{-i \epsilon}
-\frac{q_2 \sigma^2 e^{-i \epsilon} + q_3  \sigma^3 e^{-i \epsilon} }{\sqrt{2\alpha^2 - 2\alpha +1} }\Big)\nonumber
\end{eqnarray}
Here ${\bf q} = {\bf p} - {\bf K}^\pm$.
One can see that for $0<\alpha \le 1$ the values of the mentioned above topological invariants for those two Weyl points are:
$$
K^+: N_3 = N_3^{(3)} = \rev{-}1
$$
and
$$
K^-: N_3 = - N_3^{(3)} = \rev{-}1
$$
Both Weyl points are \rev{right} - handed. But one of them is of the type of "particles" while the other is of the type of the "anti - particles", i.e. the anti - Weyl point. At $\alpha = 1$ we deal with the Fermi points separated in momentum space by $\pi$. When $\alpha$ approaches to zero the two Weyl points approach each other and merge for $\alpha = 0$ at the position of $K^-$. This is the marginal Weyl point with $N_3 = \rev{-} 2$ and $N_3^{(3)} = 0$. In its small vicinity we have:
\begin{eqnarray}
K^- & : & \hat{Q}(p_0, {\bf p}) \nonumber\\ &=&   p_0 q_1
+  \frac{1}{2}(q^2_1 + p_0^2)  \, \sigma^1 e^{-i \epsilon}
-q_2 \sigma^2 e^{-i \epsilon} - q_3  \sigma^3 e^{-i \epsilon} \nonumber
\end{eqnarray}
One can calculate directly the values of $N_3$ and $N_3^{(3)}$ for this Weyl point using machinery developed in Appendix C of \cite{Zubkov:2016mvv}.

\section{Non - homogeneous and interacting systems}
\label{sectnonhom}


  In the systems with weak inhomogeneity \footnote{In case of lattice models we require that all fields depending on coordinates almost do not vary at the distances of the order of lattice spacing. Under these conditions in the majority of expressions the sum over the lattice points may be replaced by an integral.  } we come to  partition function (for $N_3,N_3^{(3)} = \pm 1$)
\begin{equation}
Z = \int D \Psi D\bar{\Psi} D e^i_k D B_k e^{i S[e^j_a, B_j,
\bar{\Psi},\Psi]}\label{ZI}
\end{equation}
with
\begin{eqnarray}
S &=&   S_0[e, B]  \nonumber\\&& + \frac{1}{2} \Bigl(i \,\int d t \,  \sum_{{\bf x}}
\bar{\Psi}_{{\bf x}}(t) e\, e_a^j  \hat \sigma^a   \overrightarrow{\hat D}_j e^{-i \epsilon} \Psi_{\bf x}(t)  \nonumber\\&& +
 \bar{\Psi}_{{\bf x}}(t) \overleftarrow{\hat D}^+_j e\, e_a^j  \hat \sigma^a    e^{-i \epsilon}\Psi_{\bf x}(t)\Bigr),\label{Se0}
\end{eqnarray}
where the sum is over $a,j = 0,1,2,3$ while $\sigma^0 \equiv 1$, and $\hat
D$ is the covariant derivative that includes the $U(1)$ gauge field $B$.
$S_0[e, B]$ is the part of the effective action that depends on $e$ and $B$
only. Now both fields $e$ and $B$ depend on coordinates. We denote here
 $$
\hat{D}^+_0 = -\partial_t e^{i\epsilon } - i \mu, \quad \hat{D}^+_k = -\partial_k + i B_k
$$
One can see, that this is the Hermitian conjugation of $i D$ except for the factor $e^{i\epsilon}$.
 In principle, one may include into covariant derivative $D$ the spin connection. However, its appearance, may be taken into account effectively by a renormalization of $e$ and $B$. Namely, the spin connection may be represented as $C_{a,k}\sigma^a$, where $C_{a, k}$ is complex - valued, and $a,k = 0,1,2,3$. Then we come to
\begin{eqnarray}
&& \bar{\Psi} e^k_b C_{a,k}\sigma^b\sigma^a \Psi + (h.c.) \nonumber\\ && = \bar{\Psi}e^k_b (C_{b,k} + i C_{a,k} \epsilon^{bad} \sigma^d)\Psi + (h.c.) \nonumber\\ && = 2\bar{\Psi}(e^k_b {\rm Re}\, C_{b,k} - e^k_b {\rm Im}\, C_{a,k} \epsilon^{bad} \sigma^d)\Psi
\end{eqnarray}
Here $(h.c.)$ means the hermitian conjugation complemented by  transformation $\Psi \to \bar{\Psi}$.
One can check that combinations  $e^k_b {\rm Re}\, C_{b,k}$ and $ e^k_b {\rm Im}\, C_{a,k} \epsilon^{bad} $ may be absorbed by redefinition of emergent gauge field.

As above one may define the two topological invariants. Let us introduce the Green function
$$
 \hat{G} = \frac{2}{e\,e_a^j  \hat \sigma^a  i \hat{D}_j +i \hat{D}_j e\,e_a^j  \hat \sigma^a}
 $$
Next, we define its Wigner transformation $G^{(M)}_{W}(p,x) = \int d^4 r e^{i\rev{(p_0r_0-{\bf p}{\bf r})}} \langle x+r/2 | \hat{G} | x- r/2\rangle   $.

After Wick rotation we introduce ${p}_0 = i \omega = i P_4$, and $P_j = p_j$ for $j = 1,2,3$; $x_0 =\rev{-} i X_4$, $X_j = x_j$, and denote the Euclidean Wigner transformation of Green function: $G_W(P,X) = G_W^{(M)}(p,x)$. We also define $Q_W$ that obeys $Q_W \ast G_W = 1$. Here by $\ast$ we denote the Moyal product
$$
\ast = e^{\frac{i}{2}(\overleftarrow{\partial_{X^i}}\overrightarrow{\partial_{P_i}}-
\overleftarrow{\partial_{P_i}}\overrightarrow{\partial_{X^i}} )}
$$
The first topological invariant is given by
\begin{eqnarray}
&& N_3  =     \frac{1}{3!\,4\pi^2\, |{\bf V}|}\int_\Sigma \,\int d^3 X  \, {\rm Tr}  \, \Big[  {G}_W(P,X) \ast d {Q}_W(P,X) \nonumber\\&& \ast  \wedge d {G}_W(P,X) \ast \wedge  d {Q}_W(\omega, P,X) \Big]\label{N3}
\end{eqnarray}
$|{\bf V}| $ is the three - volume of the system.
Here $\Sigma$ is the three - dimensional hypersurface surrounding the given singularity ${\cal M}^{(i)}$ of expression standing inside the integral. This expression resembles the one of \cite{SZ2020}. The general procedure for the construction of such invariants has been proposed in \cite{ZW2019}. In \cite{ZZ2019} such topological invariant has been considered for the interacting QHE systems.

In addition, we define
 $$
 Q_{H,W} = i\omega - Q_{W}(P,X)\Big|_{\omega = 0}
 $$
 and
 $G_{H,W}$ obeys
 $$
 Q_{H,W}\ast G_{H,W}=1
 $$
Then the second topological invariant can be defined as
\begin{eqnarray}
&& N_3^{(3)}  =     \frac{1}{3!\,4\pi^2\,|{\bf V}|}\int_\Sigma \, \int d^3 X  \, {\rm Tr}  \, \Big[  {G}_{H,W}(P,X) \ast  \label{N_H}\\&& d {Q}_{H,W}(P,X) \ast  \wedge d {G}_{H,W}(P,X) \ast \wedge  d {Q}_{H,W}(\omega, P,X) \Big]\nonumber
\end{eqnarray}
Again, $N_3^{(3)}$ and $N_3$ may differ from each other due to the nontrivial components of the emergent vierbein $e_a^0$. For the weakly dependent on coordinates $e^k_a(X)$ and $B$ (such that ${\rm sign}\,{\rm det}_{4\times 4} \, e^k_a$ and ${\rm sign}\,{\rm det}_{3\times 3} \, e^k_a$ do not depend on $X$) we still have
$$
N_3 = \rev{-}{\rm sign}\, {\rm det}_{4\times 4} \, e_a^j
$$
and
$$
N_3^{(3)} = \rev{-}{\rm sign}\, {\rm det}_{3\times 3}\, e_a^j
$$

Those two invariants provide topological classification of vierbeins for the non - homogeneous case. In the presence of interactions we are to use the complete interacting Green function $\hat{G}$:
 \begin{eqnarray}
\langle x | {\hat G}|y\rangle &=&  \frac{-i}{Z}   \int D \Psi D\bar{\Psi} D e^i_k D B_k e^{i S[e^j_a, B_j,
\bar{\Psi},\Psi]}\nonumber\\&&  \Psi(x) \bar{\Psi}(y)
\end{eqnarray}
 instead of the non - interacting one in Eqs. (\ref{N3}) and (\ref{N_H}).
 Moreover, we may use equivalently the Green function of the original multi - fermion system
  \begin{eqnarray}
&& \langle x_1, t_1 | {\hat G}|x_2, t_2\rangle =  \frac{-i}{Z}   \int D \psi D\bar{\psi} D\Phi {\rm exp}\Bigl(i R[\Phi]  \label{Gc2} \\&& + i \int d t
\sum_{{\bf x}} \bar{\psi}_{\bf x}(t) (i\partial_t - \hat{H}(\Phi)e^{-i \epsilon}) \psi_{\bf
x}(t) \Bigr) \psi_{x_1}(t_1) \bar{\psi}_{x_2}(t_2)\nonumber
\end{eqnarray}
 and its Wigner transform.

 This will give us the modified expressions for the above topological invariants.
 $\Sigma$ of these expressions is the hypersurface in phase space $(P,X)$ that surrounds a singularity of expression standing inside the integral. Positions ${\cal M}^{(i)}$ of distinct singularities play the role of the Fermi surfaces/Fermi points in case of the non - homogeneous systems. Each ${\cal M}_i$ is reduced to a Fermi point for the case of homogeneous system. $N_3$ and $N_3^{(3)}$ defined via integrals of Eq. (\ref{N3}) and Eq. (\ref{N_H}) with $\Sigma$ surrounding ${\cal M}^{(i)}$ are responsible for the topological stability of the given singularities. Only the singular transformation that leads to a change of the values of $N_3$ and $N_3^{(3)}$ may bring the singularity of one type to the singularity of another type. Correspondingly, in the cases $N_3=\pm 1$ and $N_3^{(3)} = \pm 1$ we speak of the Weyl fermions existing in the presence of emergent gauge field and emergent vierbein.

 As for the case of the homogeneous case we may prove that $\sum_{{\cal M}^{(i)}} N_3 = 0$ and $\sum_{{\cal M}^{(i)}} N^{(3)}_3 = 0$ if the corresponding Euclidean momentum space is compact. This is always true in lattice regularized models, when the axis of imaginary time is discretized as well as the coordinate space. In this case in the non - homogeneous case the number of the left - handed fermions coincides with the number of the right - handed fermions. In condensed matter systems with noncompact axis of $\omega$ this topological theorem may be broken partially as it was explained above, and $\sum_{{\cal M}^{(i)}} N_3 \ne 0$ while $\sum_{{\cal M}^{(i)}} N^{(3)}_3 = 0$.

\section{Conclusions}
\label{Discussion}

In this paper we discuss topological classification of emergent Weyl fermions in multi - fermion systems. In Euclidean space - time there would be only one topological invariant responsible for stability of Fermi points. This is $N_3$ of Eq. (\ref{rez3}) or Eq. (\ref{N3}). Correspondingly, there exist two types of low energy emergent Weyl spinor fields with minimal values of $N_3$ - those that describe left - handed and right - handed particles/antiparticles. The low energy subsystem (of the given multi - fermion system) incident in a vicinity of the given right  - handed Weyl point cannot be transformed continuously to that of the left - handed Weyl point.

In the real - time dynamics the operator $\hat{Q}$ (inverse of the Green function)  is Hermitian up to an infinitely small correction that points out the way the poles are treated in Feynmann diagrams \footnote{More precisely, operator $\hat{Q}$ is Hermitian, but its inverse $\hat{G}$ is considered in space of generalized (rather than ordinary) operator - valued functions, and the mentioned would be infinitely small correction to $\hat{Q}$ actually has the meaning of the proper definition of $\hat{Q}^{-1}$. In space of ordinary operators the inverse to $\hat{Q}$ does not exist.}. Now we are able to define the two distinct topological invariants composed of $\hat{Q}$ and $\hat{G}$. The second invariant is that of Eq. (\ref{rez2}) or Eq. (\ref{N_H}). In the class of Hermitian operators we cannot transform smoothly those $\hat{Q}$ with different values of $N_3^{(3)}$ to each other.

We come to an unexpected conclusion: in general case in multi - fermion systems with minimal values of $N_3, N_3^{(3)} = \pm 1$ the emergent Weyl fermions appear in four rather than two topologically different classes. We feel this appropriate to divide the Weyl fermions into the left and the right - handed ones, and also into the fields defining "particles" and "anti - particles", or into the Weyl points and anti - Weyl points. Without interactions we meet the fields of "particles", i.e. Weyl points only. Then the antiparticles are understood as the holes, i.e. the absence of particles in the set of occupied states.

In the presence of interactions the sign of emergent $e^0_0$ component may become negative, and the Weyl fermion of a marginal type with $N_3 = - N_3^{(3)}$ appears. We feel this instructive to call the field describing the corresponding Weyl fermion the field of "the anti - particle" type, and the Fermi point itself may be referred to as the anti - Weyl point. This is natural because charge conjugation being applied to the ordinary (left or right - handed) Weyl fermion (with $N_3 = N_3^{(3)}$) results in the Weyl fermion with $N_3 = - N_3^{(3)}$, i.e. in the Weyl fermion of the type of "anti - particle" existing in a vicinity of anti - Weyl point.

The appearance of anti -  Weyl points with $N_3 = - N_3^{(3)}$ has sense only in the presence of the conventional Weyl fermions with $N_3 = N_3^{(3)}$. Without the latter R transformation $\Psi \to - \Psi, \bar{\Psi}\to \bar{\Psi}$ brings Weyl fermions to the type of "particles" with $N_3 = N_3^{(3)}$. At the same time if both types of Weyl fermions co - exist, the interesting phenomena may occur. For example, a couple of Weyl fermions $(N_3, N_3^{(3)}) = (\rev{-}1,-1)$ and $(\rev{-}1,1)$ may merge giving marginal Weyl point with $(N_3, N_3^{(3)}) = (\rev{-}2,0)$. We gave an
example of the lattice condensed matter system, in which two \rev{right} - handed Weyl points exist. One of them is of the type of a Weyl point while another one is of the type of the anti - Weyl point. Changing smoothly parameter $\alpha$ of the system it is possible to bring it to the state, in which the Weyl point and the anti - Weyl point merge giving the marginal Weyl point with   $N_3=\rev{-}2$ and $N_3^{(3)} =0$. (The way to calculate the values of $N_3$ and $N_3^{(3)}$ for this Fermi point may be read off from Appendix C of \cite{Zubkov:2016mvv}.) Notice, that if imaginary time axis is discretized as well as coordinate space, the sum over the Weyl points of both $N_3$ and $N_3^{(3)}$ is equal to zero. As a result, in such systems the total number or Weyl points is equal to the total number of anti - Weyl points, while the total number of left - handed Weyl/anti - Weyl points is equal to the total number of right - handed Weyl/anti - Weyl points. This theorem is broken partially in the mentioned toy model, in which the imaginary time axis is not discretized, and the axis of $\omega$ remains open.

In practise the anti - Weyl points may appear in solids in the presence of sufficiently strong inter - electron interactions. Therefore, prediction of such materials cannot be given on the basis of DFT calculations only. We need for the purpose of engineering of such materials (in addition to the DFT calculation of energy bands) the methods that take into account interactions between Bloch electrons. One may also suppose that existence of anti - Weyl points may be found in fermionic superfluids, possibly, under specific external conditions.

It is worth mentioning that in addition to the topological classification of Fermi  points presented here there should exist the complimentary topological classification of the systems, in which the energy bands cross each other at the points in momentum space, but when those Weyl points form various forms of Fermi surfaces rather than the Fermi points. In the case when the values of $N_3$ (and $N_3^{(3)}$) are minimal $=\pm 1$ such a classification has been presented in \cite{Nissinen:2017ksz}. In addition to the type I Weyl points there exist the type II resulting in the Fermi surface rather than a single Fermi point, and the type III, in which energy dispersion near the Weyl point becomes degenerate. The type IV combines properties of the type II and III.  The extension of this classification to the case of non - minimal values of $N_3$ and $N_3^{(3)}$ is also possible, but it is out of the scope of the present paper.

Finally, we would like to notice that the classification presented here may be relevant for the high energy physics and applications of quantum field theory to cosmology (see \cite{Volovik2020}, \cite{Vergeles2019} and references therein). Then the appearance of the four (rather than two) topologically distinct types of Weyl fermions may be assumed from the very beginning. Such a construction may be relevant for the proper theory of quantum gravity, which should include the strong fluctuations of vierbein giving rise to all four types of the Weyl points \cite{Vergeles2019}.

The author is grateful to G.E.Volovik for useful discussions.

\end{document}